\documentclass{iau}

\usepackage{amsmath}
\usepackage{graphicx}
\usepackage{multirow}

\begin{document}

\lefttitle{Hyosun Kim}
\righttitle{AGB-pPN Evolution: Whorled Patterns and Stellar Companions}

\journaltitle{Planetary Nebulae: a Universal Toolbox in the Era of Precision Astrophysics}
\jnlDoiYr{2023}
\doival{10.1017/xxxxx}
\volno{384}

\aopheadtitle{Proceedings IAU Symposium}
\editors{O. De Marco, A. Zijlstra, R. Szczerba, eds.}
 
\title{Evolution From Asymptotic Giant Branch to Pre-planetary Nebula:
  Whorled Patterns and Stellar Companions}

\author{Hyosun Kim}
\affiliation{Korea Astronomy and Space Science Institute, 776,
  Daedeokdae-ro, Yuseong-gu, Daejeon 34055, Republic of Korea}

\begin{abstract}
Bipolar or multipolar lobes in pre-planetary nebulae (pPNe) often exhibit intertwined outer whorled patterns, resulting from stellar wind matter accumulation during the asymptotic giant branch (AGB) phase. These structures are likely triggered by stellar or substellar companions. We regard that CW Leonis currently stands at a critical transition moment, providing a vivid illustration of the progression from an AGB star in a binary system to a pPN. We have found that CW Leonis has shown significant enhancements in its optical and near-infrared light curves over the past two decades, with the recent Hubble Space Telescope image finally revealing the long-awaited central star. Utilizing an eccentric-orbit binary model, we can reproduce the position-angle dependence of the expansion velocity in the whorled pattern around CW Leonis, suggesting a nearly face-on orbital inclination. Its contradiction to the features in the innermost circumstellar envelope, corresponding to a nearly edge-on inclination, may imply the presence of an additional companion. Our updated theoretical framework explores the complexity of the whorled pattern. Further identifying and monitoring phase-transition candidates at the tip of the AGB will provide valuable insights into the AGB-pPN transition and the role of companions in shaping the morphological evolution of these stellar objects.
\end{abstract}


\maketitle

\section{Introduction}

\begin{figure} 
  \includegraphics[width=\textwidth]{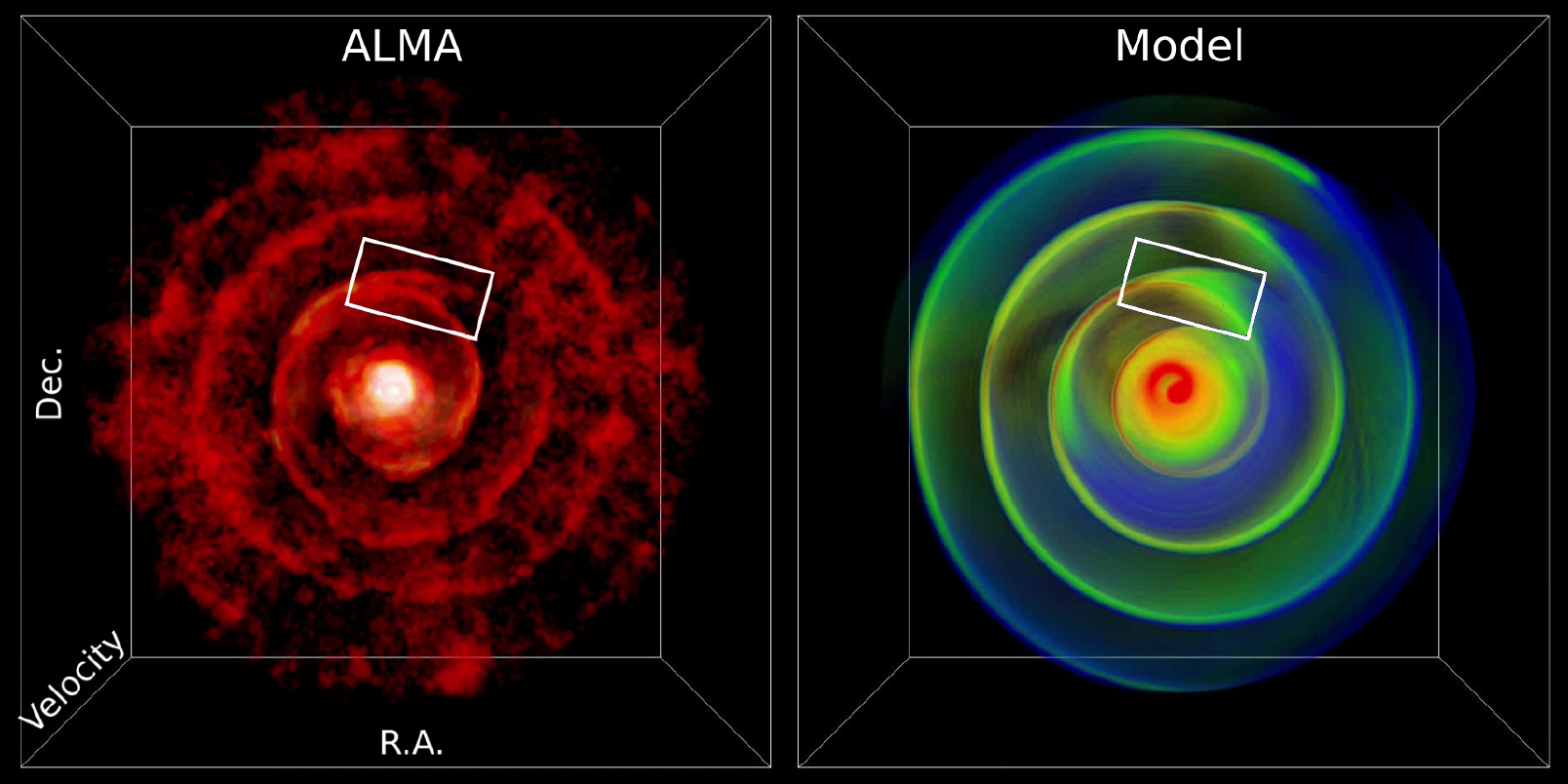}
  \caption{%
    Comparison between the Atacama Large Millimeter/submillimeter Array
    (ALMA) map of AFGL 3068 (LL Pegasi) for the $^{12}$CO molecular line
    emission and a hydrodynamic model; a snapshot of the supplementary
    animation of \citet{kim17}. Bifurcation is identified as evidence
    for an eccentric orbit of the central binary (white box).
  }\label{fig:LLp}
\end{figure}

The motivation of this work is to better understand the morphological transition between the asymptotic giant branch (AGB) and pre-planetary nebula (pPN) phases. Most circumstellar envelopes of AGB stars are spherical, or near-spherical, but the majority of (p)PNe are not spherical \citep[e.g.,][]{zuc86,sah07,sah11}. In particular, a bipolar (p)PN is believed to be formed in the condition that the matter ejection is confined by a surrounding flat disk, which is believed to be produced in a close binary star system.

As an example, V Hydrae observed with the Hubble Space Telescope indicates the high-speed bullet-type ejections every 8.5 yr, and it is proposed that the pericenter passages of an unseen companion having a very elongated orbit with such an orbital period may provoke that phenomenon \citep{sah16}. However, whether such an eccentric orbit can survive against the tendency toward circularization can be an issue \citep{sal19}.

Another example can be the PN IRAS 17150-3224. A successful modeling by \citet{hua20} showed that at least 3 different directional bipolar structures ejected at different times are overlaid in this source, in addition to the whorled pattern in the outer circumstellar envelope. The adopted geometry was attributed to the precession of a hypothesized disk, and then the disk precession was concluded to be quite severe in the angle change.

To understand the AGB-(p)PN shape transition, the important thing we need to understand is the binary system, in particular the accretion disk, at the phase \emph{immediately before} the phase transition. We consider the closest carbon star, CW Leonis (a.k.a. IRC+10216), having experiencing extraordinary brightening for nearly two decades until now, as an appropriate target so that a continuous monitoring is desired.

A simultaneously important task is to build up a theoretical framework to comprehend the stellar orbital properties and accretion processes by analyzing density and velocity distributions of the pattern whorled around the central stars.
Following the archetypal works on the spiral-shell pattern driven by reflex motion of the mass-losing star in a circular orbit \citep{sok94,mas99,kim12b}, eccentric-orbit binary cases are also intensively studied \citep[e.g.,][]{he07,kim19}. Many interferometric observations indeed have shown the whorled patterns in the circumstellar media of AGB stars, which are well reproduced by theoretical models based on the reflex motions \citep[e.g.,][see also Figure\,\ref{fig:LLp}]{mae12,kim12c,kim13,cer15,kim17,dec20,buj21}. 

This paper reports recent progress in both theoretical and observational aspects of such investigations on the whorled patterns and stellar companions, and in particular for an example carbon star CW Leonis, which is possibly currently evolving off the AGB.

\section{Propagation of Whorled Pattern Hinting at Geometry}

The propagation speed (differential proper motion) of a circumstellar whorled pattern is frequently adopted as the (often-assumed isotropic) expansion velocity of the wind.
For example, \citet{bal12} measured the angular translation of the ensemble of all observed arcs of the pPN CRL 2688 between the Hubble Space Telescope images in 2002 and 2009 epochs, from which their attempt to match this transverse velocity to the line-of-sight velocity (Doppler shift velocity measured from molecular line observations) yielded a distance of 340 pc, differing from a distance of 420 pc measured by a different method \citep{uet06}. Here, we can naturally question whether the isotropic wind assumption is valid or not.
For the same source, \citet{uet13} found that the derived transverse wind velocity varies across the Hubble Space Telescope image, but the reason for the position-angle dependence was not provided.
The transverse velocities of two AGB stars (AFGL 3068 and CW Leonis) and two PNe (NGC 6543 and NGC 7027) also show a very large dependence of the transverse velocities on the position angle \citep{gue20,kim21}.

\begin{figure} 
  \includegraphics[width=\textwidth]{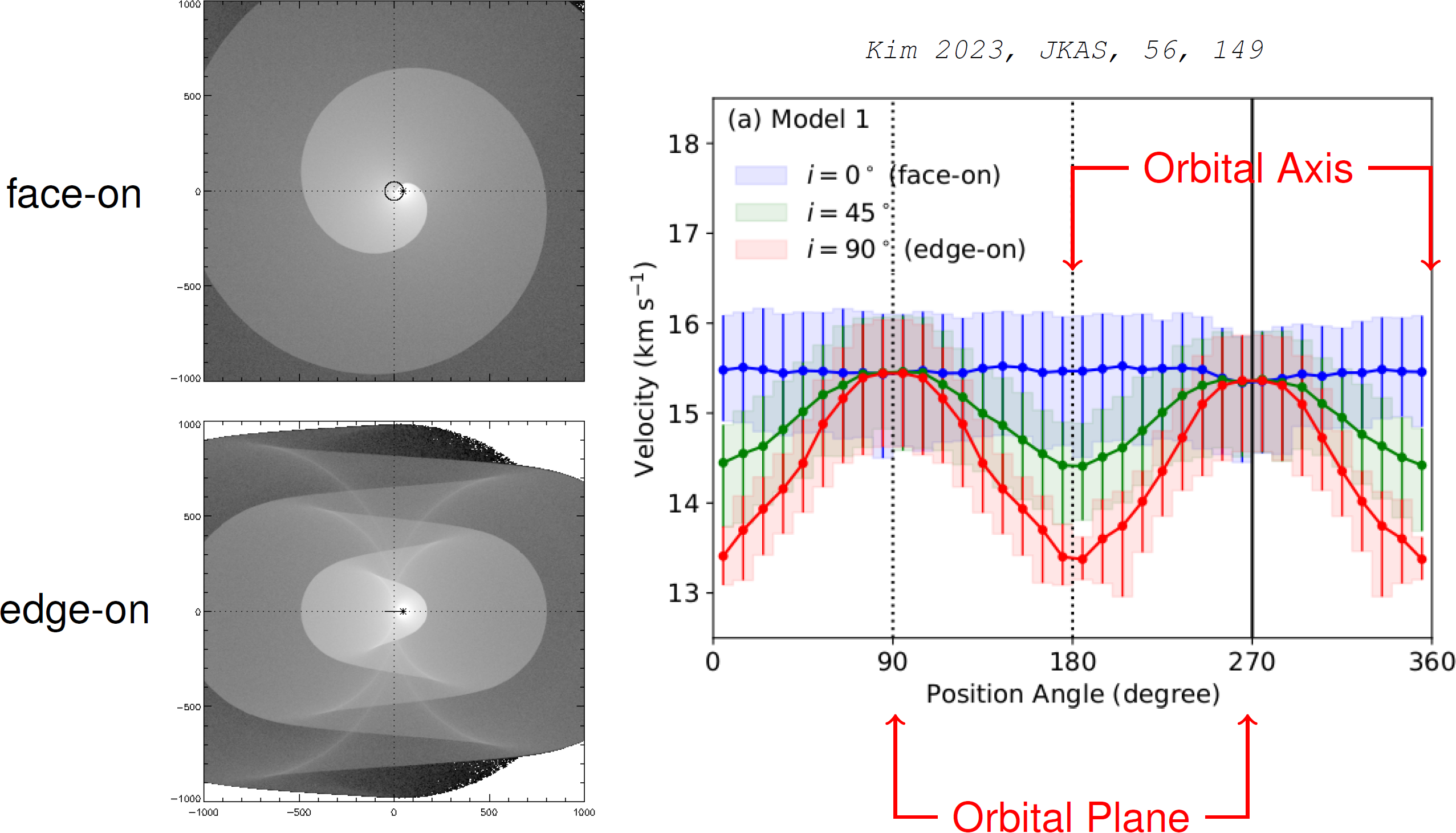}
  \caption{%
    (Left) Face-on and edge-on views of the circumstellar spiral-shell
    pattern driven by the orbital motions of the central binary system,
    calculated by a particle simulation using the first version of the
    pinwheel code \citep{kim24}. A perfectly circular orbit is assumed.
    In the bottom panel (for the edge-on image), the orbital axis lies 
    along the vertical axis.
    (Right) Figure 2(a) of \citet{kim23}. The expansion velocity profiles
    as a function of position angle at three different inclination angles.
  }\label{fig:cir}
\end{figure}

\begin{figure} 
  \includegraphics[width=\textwidth]{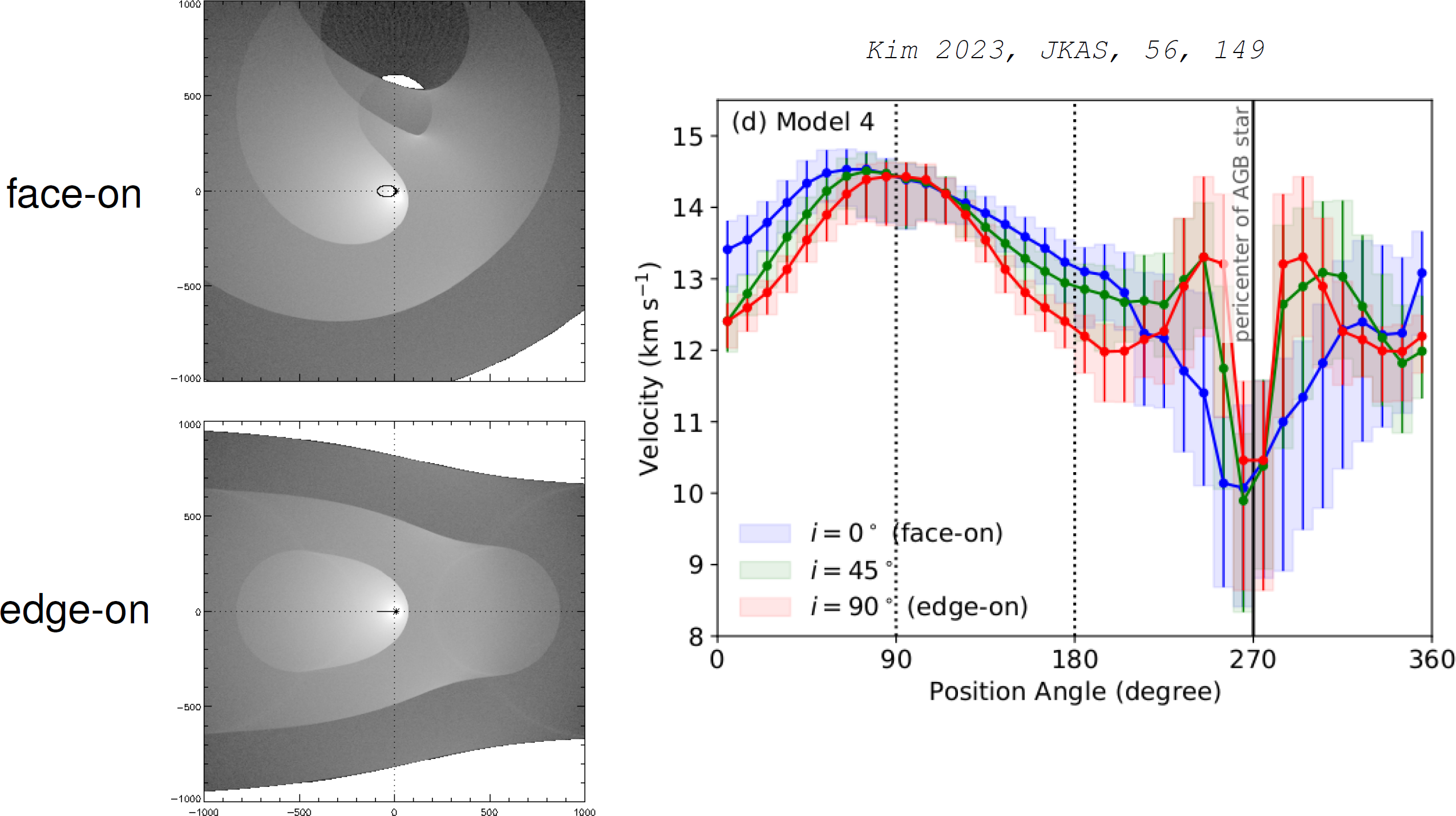}
  \caption{%
    Same as in Figure\,\ref{fig:cir}, but for an eccentric orbit binary.
    Right panel is a copy of Figure 2(d) of \citet{kim23}, which shows the
    minimum transverse wind velocity at the pericenter of the mass-losing
    star.
  }\label{fig:ecc}
\end{figure}

Recently, \citet{kim23} reported the theoretical study on the velocity variation of the spiral-shell pattern whorled around the mass-losing star with an intrinsically isotropic and continuous wind ejection, which becomes anisotropic due to the orbital motion in a binary system.
If the orbital shape is perfectly circular and the orbital axis is aligned with the line of sight (i.e., the orbital plane is face-on), the transverse velocity derived from the differential proper motion of the whorled pattern is independent on the position angle of the image (see Figure\,\ref{fig:cir}, blue). In the cases that the orbital axis is misaligned, the transverse velocity is minimized at the position angles corresponding to the projection of the orbital axis (see Figure\,\ref{fig:cir}, red and green).
Although the eccentric orbit cases are more complicated, one indubitable finding from Figure\,\ref{fig:ecc} is that the transverse wind velocity is minimized at the position angle to the pericenter of the mass-losing star, regardless the current orbital phase.

\begin{figure} 
  \includegraphics[width=\textwidth]{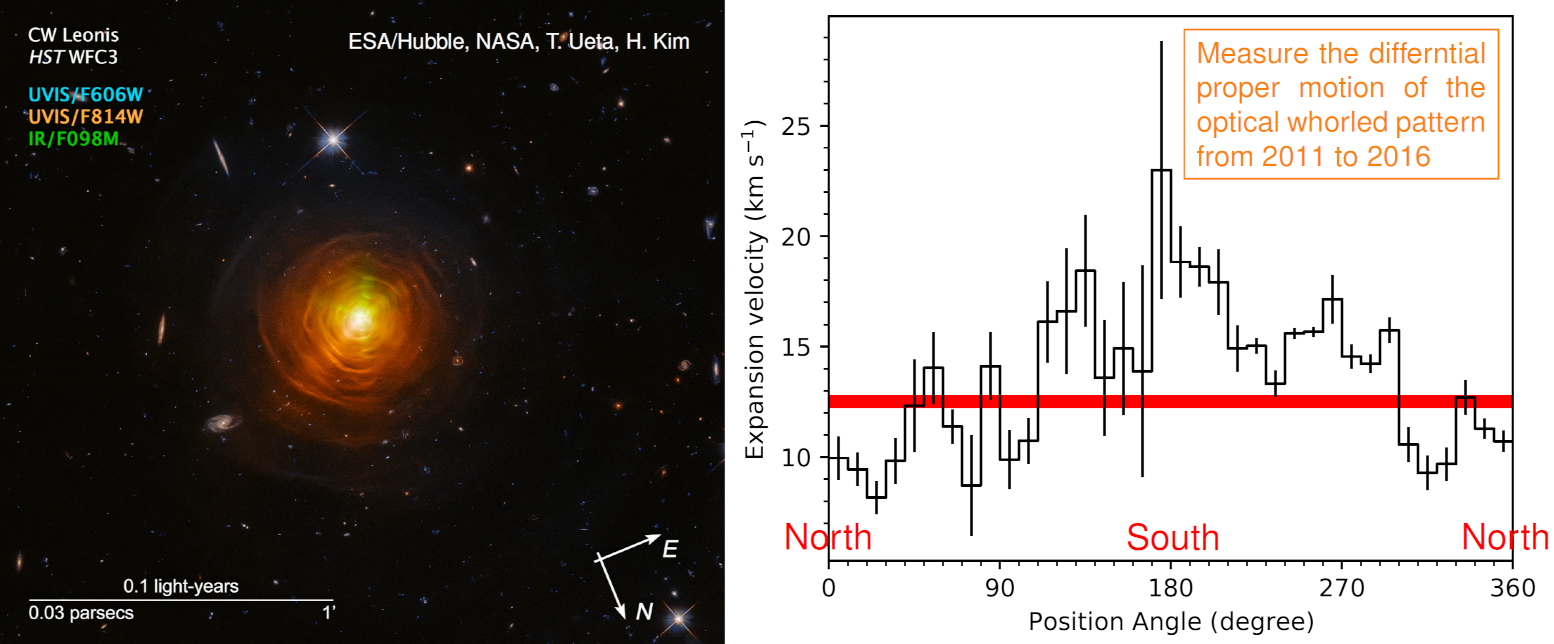}
  \caption{%
    (Left) Pseudo-color image of CW Leonis in the Hubble Space Telescope's
    gallery https://hubblesite.org/contents/news-releases/2021/news-2021-059.
    Image dimension is 2.4 arcmin across (about 0.2 light-years).
    Image credit: ESA/Hubble, NASA, Toshiya Ueta (Univ. of Denver),
    Hyosun Kim (KASI).
    (Right) Expansion velocity measured by \citet{kim21} using differential
    proper motion of the whorled pattern from the year of 2011 to 2016, as a
    function of the position angle (from north to the east).
  }\label{fig:IRC}
\end{figure}

Based on these simulational results and the position angle dependence on the expansion velocity of circumstellar pattern surrounding CW Leonis (Figure\,\ref{fig:IRC}), it is proposed that CW Leonis may be an eccentric binary with the pericenter of the AGB star projected toward the northern direction on the nearly face-on orbit \citep{kim21,kim23}.
Note that a nearly face-on eccentric-orbit binary was also suggested by \citet{gue18} to reproduce the offsets of centers of pattern shells in CW Leonis, although the pericenter located at a different position angle in this consideration.

There is, however, a problem in that the inclination (nearly face-on orbital plane) is contradictory to the previous interpretation for its geometry as nearly edge-on \citep[e.g.,][]{dec15} mainly based on the bipolar-like structure observed in the optical, being remained for at least a decade or longer until its disappearance in the 2011 Hubble Space Telescope image \citep{kim15}, and the elongation of radio continuum emission that is supposed to indicate the flattened disk-like structure, viewed edge-on, possibly paving the bipolar-like optical light path.
Here, a new suggestion can be made to engage the seemingly contradictory geometries: does CW Leonis contains a triple star system at the center having the inner companion orbit nearly edge-on and the outer companion orbit nearly face-on?

\section{Whorled Pattern in a Triple System}

\begin{figure} 
  \includegraphics[width=\textwidth]{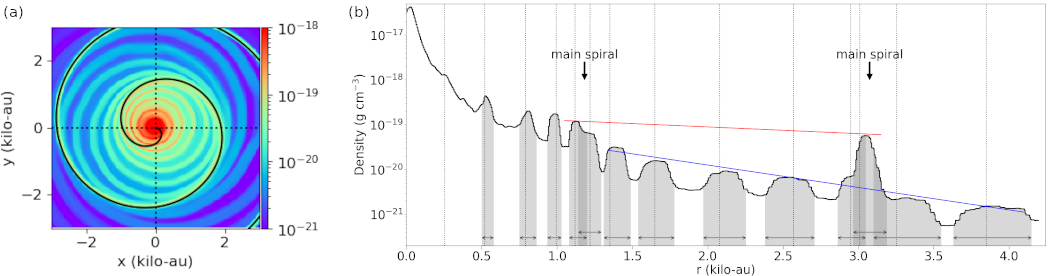}
  \caption{%
    (a) Circumstellar density pattern formed in the orbital plane of
    a triple star system using a hydrodynamic simulation, overlaid by a
    black line (main spiral, hereafter) presenting the location of the
    Archimedean spiral that would be produced by the outer binary alone.
    (b) Density profile along an angle of 45 degrees with respect to the
    $y$ axis. Individual shaded regions (and horizontal two-headed arrows)
    indicate the radial extensions of the individual fine spiral ridges,
    which converge upon each other at the main spiral. Red and blue lines
    show the difference in the radial rates of the density decrease in the
    main and fine patterns. Refer to \citet{kim24} for details.
  }\label{fig:tri}
\end{figure}

A stable, hierarchical triple system is investigated through hydrodynamic and particle simulations to provide a physical framework for interpreting the complex circumstellar patterns whorled around AGB stars \citep{kim24}. As shown in Figure\,\ref{fig:tri}, the introduction of a third star as a close companion to the mass-losing star causes a periodic variation in the orbital motion of the AGB star, creating an additional finer pattern in the otherwise inter-spiral region of the Archimedean spiral pattern that would be formed by the outer binary alone. The peak density of the finer pattern is comparable to that of the main spiral pattern in the innermost region but the density contrast of the finer pattern more quickly decreases (and plausibly becomes nondetectable) with increasing radius (blue in Figure\,\ref{fig:tri}(b)), while that of the main pattern (i.e., the Archimedean spiral) declines slowly (red in Figure\,\ref{fig:tri}(b)) by a factor that is related to the ratio between the outer and inner orbital periods. This situation may lead observers to perceive an apparent, but unreal jump in the pattern interval, as has been reported for CW Leonis \citep[e.g.,][]{mau00,gue18}. In addition, the fine spiral pattern can be very well approximated by circular rings with centers that are systemically offset from each other, which is also a known characteristic of the structure observed in the circumstellar medium of CW Leonis \citep[e.g.,][]{mau00,gue18}. The abrupt radial change in the interval of the pattern, the off-centering of the pattern, and some intertwining with other ridges that are the characteristic features found in observations of CW Leonis, all direct to this source as possibly being a triple system.

\section{Extraordinary Brightening of the Carbon Star, CW Leonis: AGB--pPN phase transition?}

\begin{figure} 
  \includegraphics[width=\textwidth]{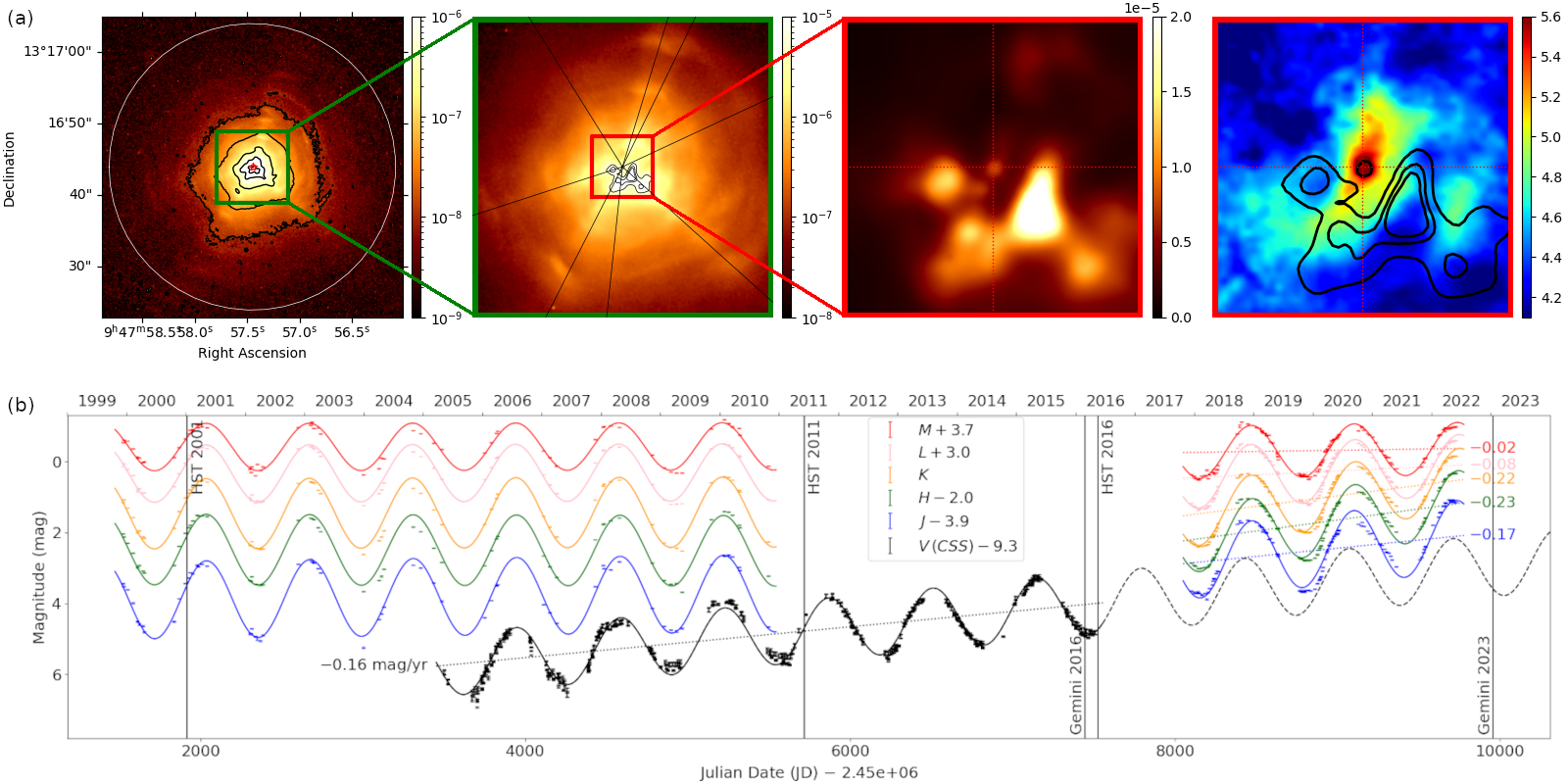}
  \caption{%
    (a) Hubble Space Telescope image at the optical 0.8\,$\mu$m band
    with the size of 20 arcsec (first panel), 10 arcsec (second panel),
    and 2 arcsec (third panel) in Jansky units; the last panel displays
    the color map (in units of magnitudes) generated by the third panel
    and the corresponding 0.6\,$\mu$m image. The reddest spot coincides
    with the proper-motion-corrected stellar position (at the junction
    of red dotted lines), where a point-like optical source is detected
    for the first time (marked by black contours of brightness). The
    radial beams and nearly circular rings appearing prominent in the
    central 5 arcsec region are the intriguing features (second panel).
    See \citet{kim21} for details.
    (b) CW Leonis' photometric data on the Catalina Sky Survey ($V$) and
    the light curve fit (black), overlaid with the $J$ (blue), $H$ (green),
    $K$ (yellow), $L$ (pink), and $M$ (red) light curves. The $JHKLM$
    data in 1999--2008 are from \citet{she11}, and the latest $JHKLM$
    data (2017--current) are obtained through private communication.
    Vertical lines indicate the previous Hubble Space Telescope epochs
    (in 2001, 2011, and 2016) and the Gemini epochs (in 2016 and 2023).
    The near- to mid-infrared light curves show gradual brightening.
  }\label{fig:cen}
\end{figure}

The extreme carbon star CW Leonis located at the tip of AGB had been completely obscured and thus invisible for at least a few decades due to the dense circumstellar matter it had ejected \citep{han98,ski98,mau00,lea06,kim15}. Therefore, it was surprising and unbelievable when a local brightness peak appeared exactly at the expected (proper-motion-corrected) stellar position in the Hubble Space Telescope's 0.8\,$\mu$m image taken in 2016 \citep{kim21} (see Figure\,\ref{fig:cen}(a), third panel). The local optical peak was compact and the reddest spot in the color map (fourth panel of Figure\,\ref{fig:cen}(a)), revealing the red nature of the evolved giant star.

Currently the star's brightness grows very fast (Figure\,\ref{fig:cen}(b)), possibly indicating that this star may be unveiling the essential clues for conditions required for the phase transition from AGB to pPN. At the same time, its recent brightening makes the adjacent clumpy structures overwhelmed (relatively dimmer). In the near-infrared, high-resolution observations using adaptive optics and speckle interferometry performed in 1995--2005 showed several near-infrared clumps \citep{ost00,tut00,wei02,ste16}, but these clumps have disappeared around 2005 \citep{ste16}. Because these previous observations were obtained with very small fields of view, therefore registering sky coordinates for these images was impossible, perhaps new monitoring observations on this source with a larger field of view, including field stars, and without central saturation may be critical to understand the late stellar evolution.

\section{Porous Envelope Scenario: Period for Redistribution of Halo Illumination linked to Binary Orbital Motion or Stellar Pulsation?}

Outer parts of the Hubble Space Telescope images of CW Leonis exhibit 8 radial beams (Figure\,\ref{fig:cen}(a), second panel). These radial beams reveal the pathways of starlight escaping from inner holes in the circumstellar envelopes, along which dust particles are illuminated.
This porous envelope scenario also explains the redistribution of relative brightness in the extended halo of CW Leonis from the elongated shape to the northwest in 2011 to the fairly symmetric shape about the central star in 2016, which
cannot be attributed to dynamical motion of the halo material \citep{kim21}. It is speculated that one of the radial beams is nearly aligned with the line of sight but with a slight angular precession directing toward the northwest in 2011 and to the line of sight (but slightly southwest) in 2016. Based on these two epochs, the precession period seems to be about two decades.
This time scale is close to the time interval of innermost shells surrounding the central star of CW Leonis, measured on an Hubble Space Telescope's optical image \citep{mau00}, which is attributed to the orbital motion of a binary.
On the other hand, these two epochs are in different stellar pulsation phases (Figure\,\ref{fig:cen}(b)), therefore the possibility of a fast rotation cannot be ruled out without more frequent monitoring observations on this source.

\section{Accretion Disk: the Key to the AGB--pPN Shape Transition}

\begin{figure} 
  \includegraphics[width=\textwidth]{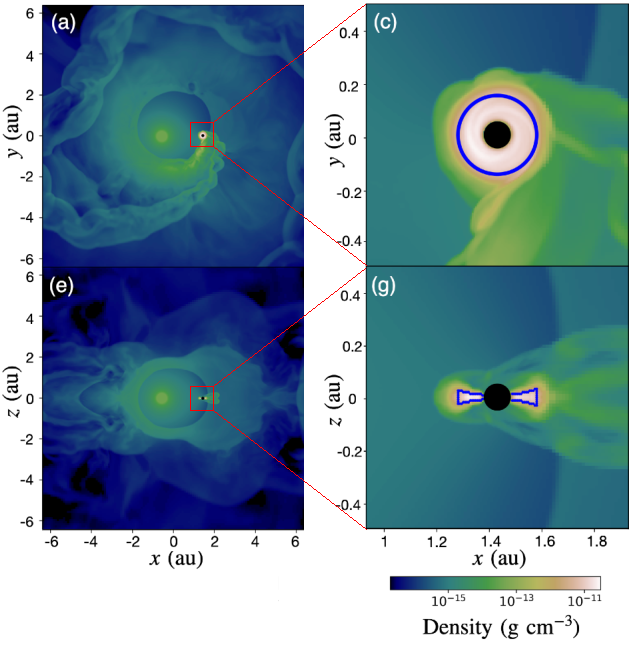}
  \caption{%
    Density distribution in a hydrodynamic simulation for the wind material
    ejected from a red giant star at $(x,\,y,\,z)=(-0.57\,\rm au,\,0,\,0)$
    circularly orbiting with its white dwarf companion separated by 2\,au.
    The right panels show the accretion disk surrounding the companion star,
    having a vertically flared shape. Blue lines indicate the surface of the
    accretion disk, as defined in \citet{lee22}.
  }\label{fig:lee}
\end{figure}

Our main motivation is to understand the physical conditions yielding the morphological transition from the AGB to the pPN, for which the accretion disk surrounding either the binary stars or the companion of the mass-losing star is one of the unavoidable part for investigation.

\begin{figure} 
  \includegraphics[width=\textwidth]{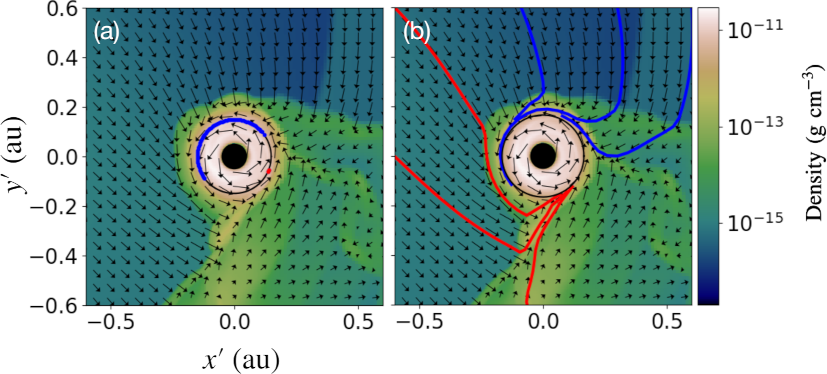}
  \caption{%
    Figure 7 of \citet{lee22}, showing the density distribution near
    the accretion disk of companion of the mass-losing red giant star, 
    overlaid by the velocity vectors of matter and some representative
    streamlines (in red and blue).
  }\label{fig:dsk}
\end{figure}

\citet{lee22} particularly focused on the inner structure of accretion disk surrounding the companion of a red giant star in a hydrodynamic simulation for a circular-orbit binary system. The disk surface, defined by a steep decline in density and temperature, indicates a vertically flared shape of the disk and two inflowing spirals in the disk are identified through close examination (Figure\,\ref{fig:lee}). By carefully tracing the matter inflowing the accretion disk, it is found that not all positions at the disk surface are passed by the streamlines; instead, the half of inflows go through one single point in the opposite side of the bow shock (red in Figure\,\ref{fig:dsk}) and the other half pass over a wide portion of the disk surface (blue in Figure\,\ref{fig:dsk}). It leads to different nature of the two spirals in the disk \citep[See][for details]{lee22}. Further investigation for the variations of the disk along the orbital phase in its size, morphology, and accretion rate would be the next step toward the comprehension on the link between the companion's disk and the late stellar evolutionary phase transition.

\section{Summary}

Whorled patterns reserve the information on the central stars, because the overall morphology of the circumstellar envelope of an AGB star is governed by the velocity fields that are influenced by the reflex motion of the mass-losing star. In addition to the orbital period and eccentricity, the inclination and the number of companions may be hinted from a careful inspection of the whorled pattern.

CW Leonis is chosen as a timely example, because its central star is getting brighter gradually and continuously for already two decades, possibly revealing the AGB-pPN phase transition. The evidences for its porous envelope are reported; the redistribution of its extended halo is interpreted as a slight angular precession of one of the radial beams, which may be also linked to the innermost pattern interval. In particular, the abrupt radial change in the interval of the pattern and the off-centering of the pattern with intertwining with other ridges suggest that CW Leonis may consist of a triple system.

Further systematic monitoring of this high mass-loss carbon star possibly at the critical transition moment by multi-scale multi-wavelength observations and multi-scale computations for theoretical investigation on the whorled patterns in multi-body systems are highly required, in order to properly approach the triggering conditions responsible for the AGB--pPN shape transition from nearly spherical to highly bipolar/multipolar morphology.
\\

H.K. thanks Victor Shenavrin for providing the latest light curves of CW Leonis. This research was supported by the National Research Foundation of Korea (NRF) grant (No. NRF-2021R1A2C1008928) and Korea Astronomy and Space Science Institute (KASI) grant (Project No. 2023-1-840-00), both funded by the Korea Government (MSIT).


\end{document}